\documentclass[11pt]{article}
\usepackage{epsfig}
\begin{document}
\setlength{\baselineskip}{0.18in}
\newcommand{\simgt}{\,\rlap{\lower 3.5 pt \hbox{$\mathchar \sim$}} \raise
1pt \hbox {$>$}\,}
\newcommand{\simlt}{\,\rlap{\lower 3.5 pt \hbox{$\mathchar \sim$}} \raise
1pt \hbox {$<$}\,}
\newcommand{\nc}{\newcommand}
\newcommand{\beq}{\begin{equation}}
\newcommand{\eeq}{\end{equation}}
\newcommand{\be}{\begin{eqnarray}}
\newcommand{\ee}{\end{eqnarray}}
\newcommand{\bi}{\bibitem}
\newcommand{\dlnk}{\partial_{{\rm ln} k}}
\def\app#1#2#3{ Astroparticle Phys. {\bf #1} (#2) #3}
\def\Aa#1#2#3{ Astron. Astrophys. {\bf #1} (#2) #3}
\def\aj#1#2#3{ Astron. J. {\bf #1} (#2) #3}
\def\apj#1#2#3{ Astrophys. J. {\bf #1} (#2) #3}
\def\araa#1#2#3{ Annu.~Rev.~Astron.~Astrophys. {\bf #1} (#2) #3}
\def\ass#1#2#3{ Astrophys. Space Sci. {\bf #1} (#2) #3}
\def\aspr#1#2#3{ Astrophys. Space Phys. Rev. {\bf #1} (#2) #3}
\def\casp#1#2#3{ Comments Astrophys. Space Phys. {\bf #1} (#2) #3}
\def\ib#1#2#3{ ibid. {\bf #1} (#2) #3}
\def\mn#1#2#3{ Mon. Not. R. Astron. Soc. {\bf #1} (#2) #3}
\def\nps#1#2#3{ Nucl. Phys. B (Proc. Suppl.) {\bf #1} (#2) #3}
\def\np#1#2#3{ Nucl. Phys. {\bf #1} (#2) #3}
\def\nat#1#2#3{ Nature {\bf #1} (#2) #3}
\def\pl#1#2#3{ Phys. Lett. {\bf #1} (#2) #3}
\def\prd#1#2#3{ Phys. Rev. D {\bf #1} (#2) #3}
\def\prep#1#2#3{ Phys. Rep. {\bf #1} (#2) #3}
\def\prl#1#2#3{ Phys. Rev. Lett. {\bf #1} (#2) #3}
\def\rmp#1#2#3{ Rev. Mod. Phys. {\bf #1} (#2) #3}
\def\zp#1#2#3{ Z. Phys. {\bf #1} (#2) #3}
\def\sjnp#1#2#3{ Sov. J. Nucl. Phys. {\bf #1} (#2) #3}
\def\jetp#1#2#3{ Sov. Phys. JETP {\bf #1} (#2) #3}
\def\jetpl#1#2#3{ JETP Lett. {\bf #1} (#2) #3}
\def\jhep#1#2#3{ J. High Energy Phys. {\bf #1} (#2) #3}
\def\ppnp#1#2#3{ Prog. Part. Nucl. Phys. {\bf #1} (#2) #3}
\def\ptp#1#2#3{ Prog. Theor. Phys. {\bf #1} (#2) #3}
\def\rpp#1#2#3{ Rep. Prog. Phys. {\bf #1} (#2) #3}
\def\yf#1#2#3{ Yad. Fiz. {\bf #1} (#2) #3}

\makeatletter
\def\alt{\mathrel{\mathpalette\vereq<}}
\def\vereq#1#2{\lower3pt\vbox{\baselineskip1.5pt \lineskip1.5pt
\ialign{$\m@th#1\hfill##\hfil$\crcr#2\crcr\sim\crcr}}}
\def\agt{\mathrel{\mathpalette\vereq>}}
\def\lsim{\raise0.3ex\hbox{$\;<$\kern-0.75em\raise-1.1ex\hbox{$\sim\;$}}}
\def\gsim{\raise0.3ex\hbox{$\;>$\kern-0.75em\raise-1.1ex\hbox{$\sim\;$}}}
\makeatother
\begin{center}
{\bf \Large
Constraints on inflation from CMB and Lyman-$\alpha$ forest
}
\bigskip
\\{\bf S.~Hannestad
\footnote{e-mail: {\tt steen@nordita.dk}}\\
{\small {\it{NORDITA, Blegdamsvej 17, DK-2100 Copenhagen, Denmark
\\
}}}}
{\bf S.~H.~Hansen \footnote{e-mail: {\tt hansen@astro.ox.ac.uk}} \\
{\small {\it{NAPL, University of Oxford, Keble road, OX1 3RH, Oxford, UK
\\
}}}}
{\bf F.~L.~Villante  \footnote{e-mail: {\tt villante@fe.infn.it}}
\\
{\small
{\it{Dipartimento di Fisica and Sezione INFN di Ferrara, Via del Paradiso 12,
44100 Ferrara, Italy\\
}}}}
{\bf A.~J.~S.~Hamilton
\footnote{e-mail: {\tt ajsh@glow.colorado.edu}}\\
{\small {\it{JILA and Dept. Astrophys. \& Planet. Sci., Box 440, U. Colorado, 
Boulder CO 80309, USA
}}}}
\end{center}

\begin{abstract}
  We constrain the spectrum of primordial curvature perturbations by
  using recent Cosmic Microwave Background (CMB) and Large Scale
  Structure (LSS) data. Specifically, we consider CMB data from the
  COBE, Boomerang and Maxima experiments, the real space galaxy power
  spectrum from the IRAS PSCz survey, and the linear matter power
  spectrum inferred from Ly$-\alpha$ forest spectra, where we for
  simplicity assume the absence of appreciable covariances.  We study
  the case of single field slow roll inflationary models, and we
  extract bounds on the scalar spectral index, $n$, the tensor to
  scalar ratio, $r$, and the running of the scalar spectral index,
  $\dlnk$, for various combinations of the observational data. We find
  that CMB data, when combined with data from Lyman$-\alpha$ forest,
  place strong constraints on the inflationary
  parameters. Specifically, we obtain $n \approx 0.9$, $r \lsim 0.3$
  and $\dlnk \approx 0$, indicating that {\it big n, big r} models
  (often referred to as hybrid models) are ruled out.
\end{abstract}


\section{Introduction}
Inflation is generally believed to provide the initial conditions for
the evolution of large scale structure (LSS) and the cosmic
microwave background radiation (CMB). The garden of inflation offers a
bounty of models, each of which predicts a certain
power spectrum of primordial curvature perturbations,
${\cal P}(k)$, a function of the wavenumber $k$.  This power
spectrum can be Taylor-expanded about some wavenumber $k_0$ and
truncated after a few terms~\cite{Lidsey:1997np} 
\be {\rm ln} {\cal P}(k)
= {\rm ln} {\cal P}(k_0) + (n-1) \, {\rm ln} \frac{k}{k_0}
+\left. \frac{1}{2} \frac{d\, n}{d \, {\rm ln}k} \right|_{k_0} \, {\rm
  ln} ^2 \frac{k}{k_0} + \cdots
\label{power}
\ee 
in which the first term is a normalization constant,
the second is a power-law approximation,
with the case $n = 1$ corresponding to a scale invariant
Harrison-Zel'dovich spectrum,
and the third term is the running of the spectral
index.

An important class of models is given by single field slow roll (SR)
inflationary models, which can be treated perturbatively. The
properties of SR models are well known (see e.g.
ref.~\cite{Lyth:1999xn} for review and a list of references), and we
will here classify the different models by the 3 parameters
$(n,r,\dlnk)$, where $n \equiv d\ln {\cal P}/d{\rm ln}k|_{k=k_0} + 1$ is
the scalar spectral index at the pivot scale $k_0$, the parameter $r$
is the tensor to scalar perturbation ratio at the quadrupole scale, and
$\dlnk \equiv dn/d{\rm ln}k|_{k=k_0}$.  The reason for using these 3
variables instead of the normal SR parameters $(\epsilon, \eta, \xi^2)$
(defined in the appendix)
is simply that the former 3 variables are more closely related to what is
measured from observations.

In SR models the tensor spectral index and its derivative can be
expressed~\cite{Lyth:1999xn,Kosowsky:1995aa}
\be n_T = - \frac{r}{\kappa} \, \, \, \, \, {\rm
  and} \, \, \, \, \, \frac{d\, n_T}{d \, {\rm ln} k} =
\frac{r}{\kappa} \left[ \left( n-1 \right) + \frac{r}{\kappa} \right]
~.
\label{consist}
\ee 
The factor $\kappa$ in the above equations depends on the model,
and in particular is different for different $\Omega_\Lambda$~\cite{knox}.
In our
analysis we use the parametrisation from ref.~\cite{Turner:1996ge},
which for the models considered here means $5 \leq \kappa \leq 7$.

The different SR models are traditionally~\cite{Dodelson:1997hr}
categorised into 3 main groups according to the relationship between
the first and second derivative of the inflaton potential, and they
are distributed in $(n,r)$ and $(n,\dlnk)$ space as in Fig.~1 (see
details in the appendix or in ref.~\cite{hhv}), where the dashed lines
on the borders between the different models are the two attractors
found in ref.~\cite{hoffman}, $r=0$ and $r=- \kappa (n-1)$.
For large
values of $n-1$ and $r$, SR models ``naturally'' predict sizeable
deviations from a power law approximation (i.e. $\dlnk\neq 0$),
so when comparing to observations it is important to include the third
parameter $\dlnk$, in addition to the parameters $n$ and $r$
commonly considered.

\begin{figure}[htb]
\begin{center}
\epsfig{file=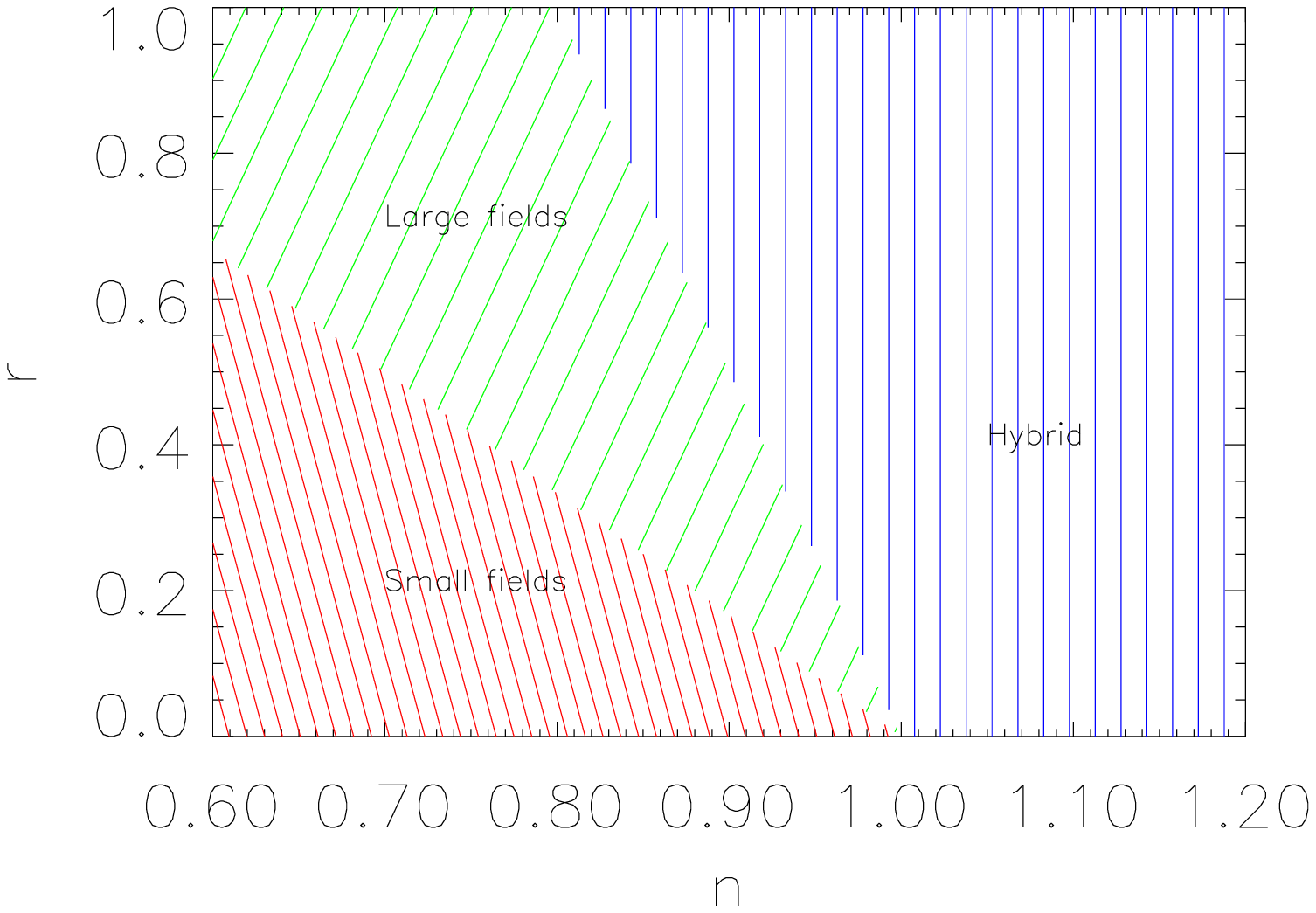,height=5cm,width=7cm}
\epsfig{file=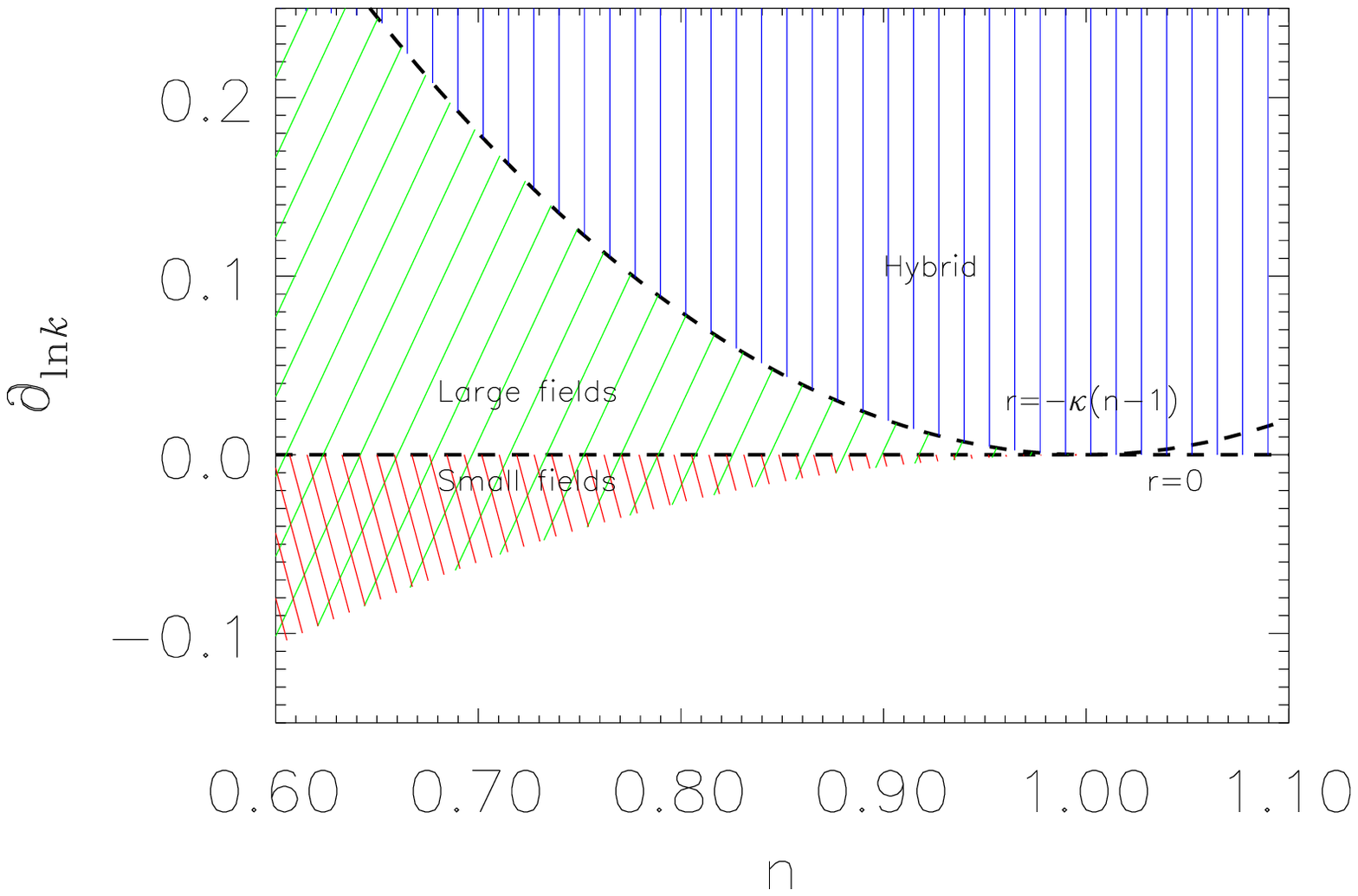,height=5cm,width=7cm}
\end{center}
\caption{The various slow-roll models in $(n,r)$ and $(n,\dlnk)$ space. 
  The dashed lines are the two attractors (here we have used $\kappa =5$
  which is a typical value for a flat universe with a large cosmological 
  constant).
  The figure $(n,\dlnk)$ is for $\xi^2=0$. The hatched regions
  move up and down by inclusion of the third derivative, $\xi^2 \neq 0$ 
  (see Eq.~\ref{appendixEQ1} in the appendix).}
\label{fig1}
\end{figure}

The purpose of this paper is to ascertain
what constraints currently available observational data
can place on the parameters $(n,\dlnk,r)$ of SR inflation.
Of course the universe
could lie outside the hatched regions of Fig.~1, indicating that one
should look beyond SR inflation;
but it seems reasonable to explore the simplest models in the first instance.
In ref.~\cite{hhv}, CMB data were
used to put bounds on inflationary parameters, and indications were
found towards a negative bend of the primordial power spectrum
($\dlnk\le 0$).
However, the scales
probed by CMB are limited, and the constraint on $\dlnk$ was weak.

In the present paper we extend the analysis of~\cite{hhv} by
including information not only from the CMB,
but also
from the linear galaxy power spectrum
measured from the IRAS Point Source Catalogue Redshift (PSCz) survey
\cite{htp},
and
from the linear matter power spectrum
inferred from the Ly$-\alpha$ forest in quasar spectra
\cite{croft}. 
Even though it is not completely clear that the correlations in the 3d
matter power spectrum are negligible, we treat them as such for
simplicity. The size of and effect from such covariances might be
essential, and should be considered more carefully in a future
investigation.
As we will show, these data, probing different ranges
of scales, yield much tighter constraints on the
inflationary parameters than CMB data alone.

\section{The data}
Recently observational data on both the CMB and LSS have improved dramatically.
The balloon-borne experiments Boomerang~\cite{boom} and
MAXIMA~\cite{max} have determined the CMB angular power spectrum
beyond the first acoustic peak, whose position near $l\sim 200$
indicates a flat universe, hence seemingly confirming the inflationary
paradigm.  The two experiments, together with COBE-DMR~\cite{COBE},
provide a high signal-to-noise determination of the CMB power spectrum
for $2 < l \lsim 800$, thus testing the structure formation paradigm on
scales roughly of the order $k\lsim 0.1$ $h$ Mpc$^{-1}$.

LSS data, probing smaller scales, provide complementary
information. In this paper we consider the real space galaxy power
spectrum from IRAS Point Source Catalogue Redshift (PSCz) survey
\cite{pscz}
which probes scales in the range $k=0.01-300$ $h$ Mpc$^{-1}$~\cite{htp,hamteg}.
In order to avoid problems with the
interpretation of non-linear effects, we use data only at scales
$k \leq 0.3$ $h$ Mpc$^{-1}$.

Further, we consider the information on the linear matter spectrum at
redshift $z\sim 3$ which can be inferred from the Ly-$\alpha$ forest
in quasar spectra.  The fact that the nonlinear scale is
smaller at higher redshift makes it possible to probe the linear
power spectrum to smaller scales ($k\sim 0.2-5$ $h$ Mpc$^{-1}$)
than are accessible to galaxy surveys at low redshift.
In this paper
we use a recent determination \cite{croft} of the linear matter power spectrum
from the Lyman-$\alpha$ forest at redshift $z=2.72$, based
on a large sample of Keck HIRES and Keck LRIS quasar spectra.
To convert the measured power spectrum of Ly-$\alpha$ flux
into the matter power spectrum, \cite{croft} apply a correction factor
obtained from $N$-body computer simulations.
To allow for possible systematic uncertainty \cite{zht}
in this correction factor,
we will repeat the analysis excluding the data at the smallest scales.

\subsection{Data analysis}
\label{dataanalysis}
In order to investigate how the CMB, PSCz and Ly-$\alpha$ data
constrain the SR parameter space $(n,r,\dlnk)$, we performed a
likelihood analysis of the data sets from COBE~\cite{COBE},
Boomerang~\cite{boom} and MAXIMA~\cite{max}, together with the
decorrelated linear power spectrum of PSCz galaxies
for $k\le0.3$ $h$ Mpc$^{-1}$~\cite{htp, hamteg}, and
the Ly-$\alpha$ data from Table~4 of ref.~\cite{croft}.
The likelihood function is
\begin{equation}
{\cal L} \propto \exp(-\chi^2/2),
\end{equation}
where 
\begin{equation}
\chi^2 = \chi^2_{\rm CMB} +  \chi^2_{\rm PSCz} +  \chi^2_{{\rm Ly-}\alpha}.
\end{equation}
For the CMB data
\begin{equation}
\chi^2_{\rm CMB} = \sum_i \frac{(C_{l,i}(\theta)-
C_{l,i})^2} {\sigma^2(C_{l,i})},
\label{eq:like}
\end{equation}
while for PSCz and Ly-$\alpha$ data
\begin{equation}
\chi^2_{{\rm PSCz, Ly-}\alpha} = \sum_i \frac{(P_{k,i}(\theta)-
P_{k,i})^2} {\sigma^2(P_{k,i})},
\label{eq:like2}
\end{equation}
the sum being taken over
published values of band-powers $C_{l,i}$ and $P_{k,i}$.
The quantity $\theta$ is a vector of cosmological parameters,
taken here to be
\begin{equation}
\theta = \{\Omega_m,\Omega_\Lambda,\Omega_b,H_0,\tau,Q,n,r,\dlnk \}~.
\end{equation}
The parameters are:
the matter density $\Omega_m$;
the baryon density $\Omega_b$;
the Hubble parameter $H_0$;
the optical depth $\tau$ to reionization;
the normalizations $Q_{\rm CMB}$, $Q_{\rm PSCz}$, and $Q_{{\rm Ly-}\alpha}$
of the CMB, LSS, and Ly-$\alpha$ power spectra;
and the inflationary parameters $n,r,\dlnk$.
We have assumed that the universe is flat,
$\Omega_\Lambda = 1-\Omega_m$,
as predicted by standard inflationary models.
For all the figures we marginalize over all other parameters.
We consider only two values for the baryon density,
$\Omega_b h^2 = 0.019$ and $0.030$, however, as we will see, the
results are very similar in those two cases, suggesting that the
results will be similar if allowing $\Omega_b h^2$ as a free parameter.

The CMB, LSS, and Ly-$\alpha$ data are all subject to uncertainties in
their overall normalizations.
The CMB groups quote estimated calibration errors
for their experiments,
which we account for by allowing the data points to
shift up or down, by 10\% for Boomerang \cite{boom},
and by 4\% for MAXIMA~\cite{max}.
Because of uncertainty in the linear galaxy-to-mass bias
for PSCz galaxies~\cite{hamteg},
we conservatively treat the normalization $Q_{\rm PSCz}$
as an unconstrained parameter.
Ref.
\cite{croft}
quotes uncertainties in the overall normalization of the matter power spectrum
inferred from the Ly-$\alpha$ forest,
but these uncertainties are based on simulations with $\dlnk = 0$,
so to avoid possible bias
we again conservatively treat the normalization $Q_{{\rm Ly}\alpha}$
as an unconstrained parameter.

For simplicity all data points have been treated as uncorrelated
in the likelihood functions, eqs.~(\ref{eq:like},\ref{eq:like2}).
For the CMB data,
correlations between estimates $C_{l,i}$ of angular power
at different harmonics $l$ are induced by finite sky coverage,
but in practice the CMB teams quote band-powers at sufficiently well-separated
bands of $l$ that the correlations are probably small.
For the PSCz data,
the published band-powers are explicitly decorrelated.
For the Ly-$\alpha$ data,
the covariances between estimates of the flux power spectrum are small,
according to Fig.~12 of \cite{croft},
and this may translate into small statistical covariances
in the inferred matter power spectrum.
As mentioned earlier, it is not completely clear how good this translation
from flux power to matter power is, and we leave this question for future
investigation.

We have chosen the pivot scale in Eq.~(\ref{power}) as
$k_{0}=0.05\,{\rm Mpc}^{-1}$. This choice is made for convenience,
since $k_{0}=0.05\,{\rm Mpc}^{-1}$ is the scale at which wave-numbers
are normalised in the CMBFAST code. Our results are independent
of the value of $k_{0}$.

\begin{figure}[tbh]
\begin{center}
\epsfig{file=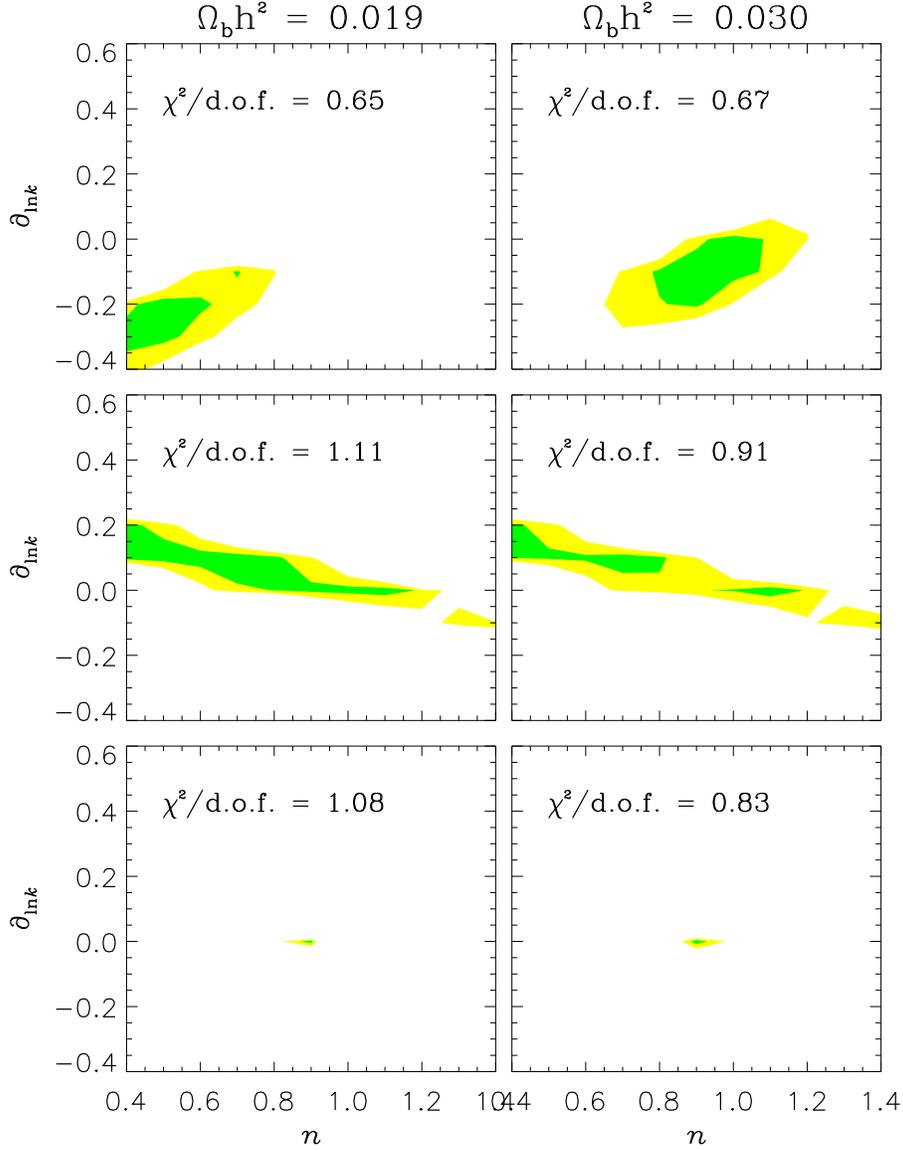,height=14cm,width=10.23cm}
\end{center}
\bigskip
\caption{The 1 and 2 $\sigma$ allowed regions for the two slow roll parameters
  $n$ and $\dlnk$.  The left panels assume a BBN prior on $\Omega_b
  h^2 = 0.019$, whereas the right panels are for $\Omega_b h^2 =
  0.030$, the value which best fits the CMB data.  The top row is for
  CMB data alone, the middle row is for Lyman-$\alpha$ data alone, and
  the bottom row is for the combined analysis.}
\label{fig2}
\end{figure}
\clearpage

\subsection{Results}
%
The analysis of the constraints from CMB data was presented in detail
in ref.~\cite{hhv} (see also~\cite{Kinney:2000nc}).  Here we extend the
analysis in \cite{hhv} by including the reionization optical depth, $\tau$, as
a free parameter. This is potentially important since there is a well-known
degeneracy between $\tau$ and the scalar spectral
index $n$.
However,
the analysis turns out to prefer models with $\tau \approx 0$,
so including $\tau$ leaves the main conclusions of \cite{hhv} essentially unchanged,
as can be seen in the upper panels of Figs.~2 and 3.
We recall here the main
conclusions from~\cite{hhv}: {\it (1)} if we allow the primordial power
spectrum to bend, $\dlnk \neq 0$, then CMB data do not constrain
the tensor to scalar perturbation ratio $r$; {\it (2)} if we assume a
BBN prior, $\Omega_b h^2 = 0.019$ \cite{tytler},
then CMB data favour a negative bend,
$\dlnk<0$, corresponding to a bump-like feature centered at scales
$k\sim 0.004\,\rm{Mpc}^{-1}$.

Let us now discuss the information on the inflationary parameters
which emerges from PSCz and Ly-$\alpha$ data.  The first observation
is that PSCz data do not provide relevant constraints in the
$(n,\dlnk)$ plane. This is easily understood, because the PSCz data at
large scales (say $k<0.03 \, h \, \rm{Mpc}^{-1}$) have large errors and
therefore play little role in the $\chi^2$
evaluation. This implies that PSCz data effectively span only one
decade in $k$.  Considering that the overall normalization of the data
is taken as a free parameter, it is evident that one cannot obtain
strong constraints from such a small range of scales.

The situation is quite different with Ly-$\alpha$ data, which have
small error bars and span almost two decades in $k$.  This is shown in
Fig.~2, where we present the allowed regions corresponding to 1 and 2
$\sigma$ 
(we define the $1\sigma$ region as $\chi^2 \leq \chi^2_{\rm min} + 2.3$
and the 2$\sigma$ region as $\chi^2 \leq \chi^2_{\rm min} + 6.17$) 
for both $\Omega_b h^2= 0.019$ as suggested by BBN (left
column), and $\Omega_b h^2= 0.03$ as suggested by CMB (right column).
The top graphs are from CMB data alone, the middle graphs are obtained
using Ly-$\alpha$ data alone, and the bottom graphs are from the
combined analysis.

It is straightforward to understand how Ly-$\alpha$ data select the allowed
regions shown in the middle panels of Fig.~2. 
At the small scales probed by Ly-$\alpha$ data, 
the theoretical linear matter power spectrum $P(k)$ 
is roughly proportional to
\begin{equation}
P(k) \propto \ln^{2}(\alpha k/\Gamma h) \, k^{n'-4}
\label{cdm}
\end{equation}
where $\alpha=2.205$, $\Gamma=\Omega_m h$ and $k$ is expressed in 
$\rm{Mpc}^{-1}$ \cite{ma}. The parameter 
$n'=d \ln {\cal P}(k_{\alpha}) /d\ln k + 1$ is the effective
spectral index of primordial density perturbation at the scale 
$k_{\alpha} \approx 3.7 (\Omega_m h^2/0.4)^{1/2}$ Mpc$^{-1}$ 
representative of Ly-$\alpha$ data
\footnote{The observational units for wavenumbers are 
$(\rm{km}\,\rm{s}^{-1})^{-1}$.
The conversion to Mpc$^{-1}$ is model-dependent since it requires the 
evaluation of the Hubble constant $H(z)$ at redshift $z=2.72$.}.
 On the other hand the Ly-$\alpha$ data, 
as discussed in \cite{croft}, are well fitted by a power law, 
$P(k)\sim k^{\nu}$, with spectral index $\nu=-2.47\pm0.06$.
This means that
\begin{equation}
\frac{d\ln P(k_\alpha)}{d \ln k} =
(n'-4) + \frac{2}{\ln(\alpha k_{\alpha}/\Gamma h)} = -2.47\pm0.06 ~.
\label{nu}
\end{equation}
If we consider that $\Omega_b h^2\lsim\Gamma h\lsim 1$, the previous
expression can be directy translated into a bound for the effective
spectral index of primordial density perturbations.  We obtain $n'\sim
0.75-1.1$, which roughly corresponds to what is shown in
the middle panels
of Fig.~2 for $\dlnk = 0$.  It is also easy to understand the
observed correlation between $n$ and $\dlnk$.  The effective spectral
index $n'$ can be expressed, as a function of $n$ and $\dlnk$,
by $n'=n+\ln(k_{\alpha}/k_0)\dlnk$. It is thus evident that the
allowed region in the plane $(n,\dlnk)$ lies roughly along lines of
constant $n'$.

It is important to note that the regions constrained by Ly-$\alpha$
data are ``orthogonal'' to the regions constrained by CMB data. This
means that the combined analysis (CMB+Ly-$\alpha$) gives much stronger
bounds than any of the two alone. Moreover, since CMB data provide a bound 
on $\Omega_m h^2$, the degeneracy  between $n'$ and 
$\Gamma h$ in Eq.~(\ref{nu}) is removed, and therefore the 
Ly-$\alpha$ observation of
$P(k)\propto k^{-2.47\pm 0.06}$ can be directly translated into a constraint
on $n'$.
This is clearly shown in the lower panels of Fig.~\ref{fig2}, 
from which one obtains the following conclusions: 
{\it (1)} the combined analysis strongly indicates that
the bend of the primordial power spectrum is close to zero, being
$-0.05 <\dlnk < 0.02$ at the 2$\sigma$ level; {\it (2)} the spectral
index $n$ is constrained to be fairly close to $n=0.9$. One notes that
the limits obtained do not crucially depend on the assumed values of
$\Omega_b$, indicating that the inclusion of Lyman-$\alpha$ data
avoids the problem with CMB data alone, that different $\Omega_b$
extend the allowed range of the spectral index beyond $n \approx
1$~\cite{Kinney:2000nc}. However, one should note that, due
essentially to CMB data, the goodness of the fit is quite sensitive to
the assumed value of $\Omega_b$, the $\chi^2$ being substantially
smaller for high $\Omega_b$.

\begin{figure}[hbt]
\begin{center}
\epsfig{file=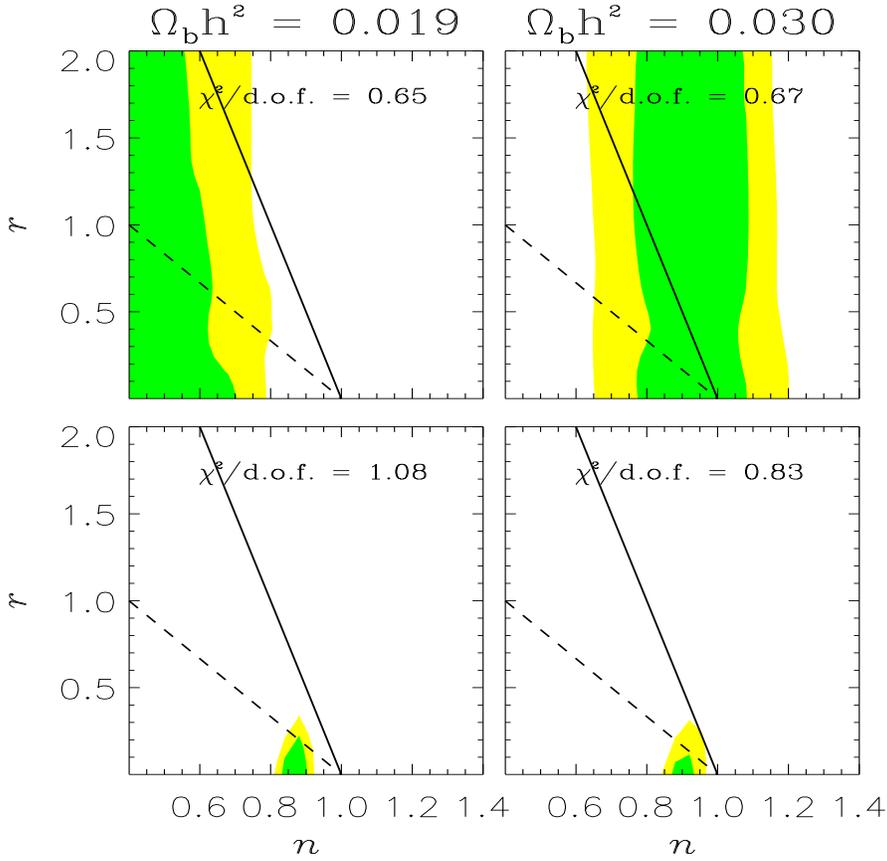,height=10cm,width=10cm}
\end{center}
\bigskip
\caption{The 1 and 2 $\sigma$ allowed regions in the $(n,r)$ plane.
  The left panels assume a BBN prior on $\Omega_b h^2 = 0.019$,
  whereas the right panels are for $\Omega_b h^2 = 0.030$, the value
  which best fits the CMB data. The top panels are for CMB data alone,
  while the bottom panels are for the combined analysis.  Hybrid
  models are to the right of the full line.}
\label{fig3}
\end{figure}

In Fig.~3 we present the constraints obtained for the remaining SR
parameter $r$. Specifically, we show the 1 and 2 $\sigma$ allowed
regions in the $(n,r)$ plane, for both low and high $\Omega_b$. The
top panels are from CMB data alone, while the bottom ones are from the
combined analysis. It is clear that in the combined analysis $r$ is
constrained to be smaller than about 0.3 at 2 $\sigma$.  This is
substantially different from the analysis of CMB data alone, where no
constraints on $r$ could be obtained, since a strong bend, e.g.
$\dlnk=-0.2$, could allow the tensor component to be big~\cite{hhv}.
It is worth noting, that Ly-$\alpha$ data don't probe $r$ directly;
they fix $n$ to be less than 1, and $\dlnk$ close to zero. In this way
strong constraint on $r$ can be obtained from CMB data.  This result
is extremely important when we compare to theoretical models, since
different classes of models predict different relationships between
the scalar spectral index $n$ and the tensor to scalar ratio $r$ (see
Fig.~1).  The combined analysis, indicating $n\approx 0.9$ and $r
\lsim 0.3$, favours small field models and seems to exclude hybrid
models. This is clear from the lower panels of Fig.~3, where the {\it
big n, big r} (often referred to as hybrid models, to the right of the
full line), are excluded at 2$\sigma$.~\footnote{We have used
$\kappa=5$ (see Eq.~\ref{consist}), and for larger $\kappa$ the hybrid
models move to bigger $r$. It is worth pointing out, that
the classification of hybrid models~\cite{linde} as in Fig. 1, and as
the models to the right of the full line in Fig. 3 is
oversimplified. In this paper we follow ref.~\cite{Dodelson:1997hr},
and by "hybrid models" we refer to potentials for which $\epsilon < \eta$
(see appendix for details).  When considering more general potentials,
or F-term hybrid inflation~\cite{linderiotto}, more complicated
behaviour results, and a simplified classification is impossible.}

An important caveat to this analysis concerns possible systematic
uncertainties in the inference of a linear matter power spectrum
from the Ly-$\alpha$ data.
What \cite{croft} measure directly from observations is the power
spectrum of transmitted Ly-$\alpha$ flux.
The conversion to a linear matter power spectrum
involves a fairly large, scale-dependent correction
which \cite{croft} extract from collisionless computer simulations,
with the Ly-$\alpha$ optical depth taken proportional
to a certain power of the dark matter density \cite{rauch}.
It has been suggested \cite{zht}
that the relation between baryonic and dark matter densities
could introduce significant uncertainty in the flux-to-mass
correction at small scales.
Specifically, pressure effects cause the baryonic density $\delta_b$
to be smoothed compared to the dark matter density $\delta_d$,
an effect that can be parametrized as
$\delta_b ({\bf k}) = \exp[-(k/k_f )^2] \delta_d ({\bf k})$,
with comoving filter scale $k_f \sim 35 \, h \, {\rm Mpc}^{-1}$
\cite{gnedin}.
The procedure considered by \cite{zht} is to treat the filter scale
$k_f$ as a free parameter constrained only by the shape of the
power spectrum of Ly-$\alpha$ flux.
However, \cite{croft} argue that treating $k_f$ as a free parameter
is overly pessimistic, and that if $k_f$ takes values
suggested by hydrodynamic simulations,
then the effect on the power spectrum is minor.

To allow for the possibility that the systematic errors are
underestimated at small scales,
we repeated the analysis neglecting the last 3
data-points from the Ly-$\alpha$ data \cite{croft},
both for $\dlnk=0$ and for $\dlnk$ free.
The upper panels in Fig.~4 show the case where $\dlnk=0$.
Here, the results are
essentially unchanged by the removal of the data points.
However, the lower panels show the full case. Here, the tight constraint
on $r$ disappears completely, and the constraint on $n$ is significantly
weakened. The reason is that the small scale data points are the
most important for constraining $\dlnk$. When these points are
removed, $\dlnk$ is not nearly as tightly constrained as before, and
as seen for the case where only CMBR data is used, a large $r$ can
be compensated by a negative bend of the spectrum.

Thus, the very tight constraint derived above depends on the
correctness of the small scale Ly-$\alpha$ data. 
Therefore, it is highly
desirable that a better understanding of the possible systematic
errors in determining matter power spectra from Ly-$\alpha$
forest observations is developed.

\begin{figure}[hbt]
\begin{center}
\epsfig{file=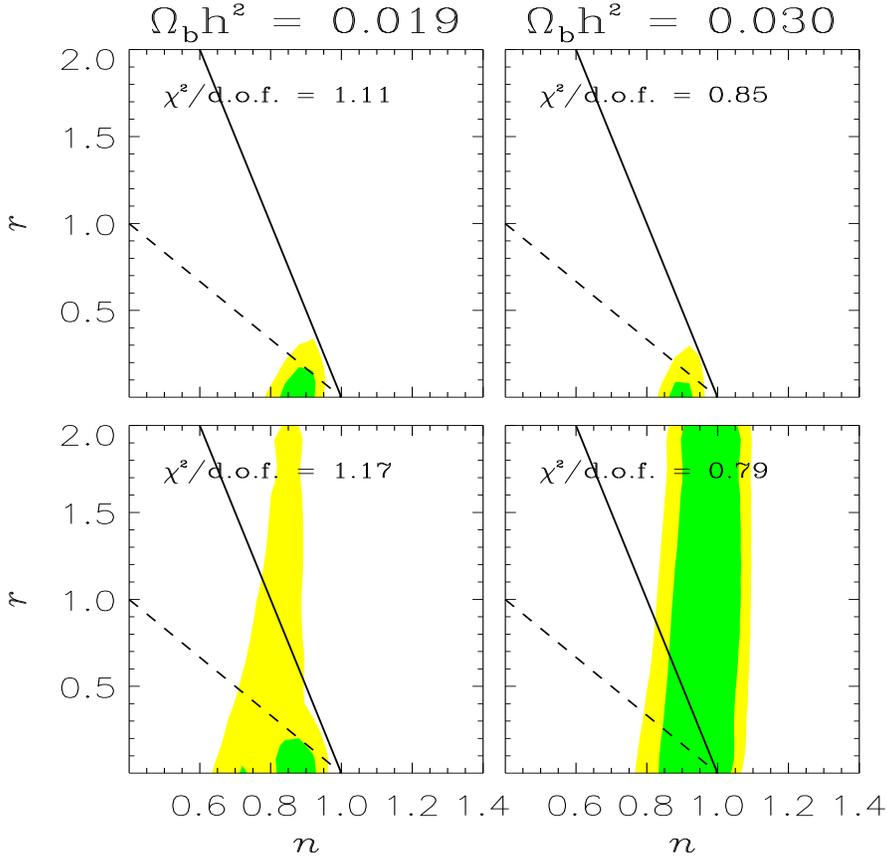,height=10cm,width=10cm}
\end{center}
\bigskip
\caption{The 1 and 2 $\sigma$ allowed regions in the $(n,r)$ plane.
  These results are obtained from the combined analysis, neglecting
  the last 3 data-points from~\protect\cite{croft}.  The left panels
  assume a BBN prior on $\Omega_b h^2 = 0.019$, whereas the right
  panels are for $\Omega_b h^2 = 0.030$, the value which best fits the
  CMB data. The top panels assume $\dlnk=0$, whereas the lower panels
  have $\dlnk$ as a free parameter. Hybrid models are to the right of
  the full line.}
\label{fig4}
\end{figure}

Another worry is that the error ellipses for CMB and Ly-$\alpha$ are
only marginally overlapping.  This might indicate that the two data
sets are mutually inconsistent, and that other physical effects should
be taken into account.  For the Ly-$\alpha$ data one might think of
massive neutrinos or warm dark matter, where both would suppress power
on small scales, potentially allowing bigger $n$ or bigger positive
$\dlnk$.

\section{Conclusions}
We have considered data from both
CMB and LSS to place constraints on the parameters of
single field slow roll inflationary models.
We have found that, by combining CMB data with the power spectrum
inferred from Lyman$-\alpha$ forest, one obtains strong
constraints on the inflationary parameters. 
We obtain $0.8\lsim n \lsim 1.0$, 
$r \lsim 0.3$ and $-0.05 \lsim \dlnk \lsim 0.02$ at $2\sigma$ level.
In the language of SR this means $\epsilon < 0.03$ and
$\eta < 0.06$, with the best fit model being small fields. This still
leaves a large part of the SR parameter space open, but seems to
exclude {\it big n, big r} models, often referred to as hybrid models.
These constraints are much stronger than those from CMB 
data alone, and arise because the
error-ellipse from Ly-$\alpha$ data is almost perpendicular to the
error-ellipse from CMB data. 
Let us repeat, that we in this analysis for simplicity have assumed
the absence of appreciable covariances between the reconstructed
Ly-$\alpha$ mass powers.

\section*{Acknowledgements}
We are pleased to thank A.~Dolgov, A. Linde, A.~Melchiorri and J.~Silk 
for comments and discussions.  
SHH is supported by a Marie Curie Fellowship of the European Community 
under the contract HPMFCT-2000-00607.
We acknowledge the use of CMBFAST~\cite{cmbfast}.

\appendix
\section{Notation}
We use the notation: $\eta\equiv \alpha \, \epsilon$, where
\[
\epsilon= \frac{M^2}{2} \left( \frac{V'}{V} \right)^2 
\, \, \, ~,\, \, \, 
\eta = M^2 \frac{V''}{V} - \frac{M^2}{2} \left( \frac{V'}{V} \right)^2
\, \, \, \mbox{and}\, \, \,
\xi^2 = M^4 \frac{V'V'''}{V^{2}}
\]
see \cite{hhv,Kinney:2000nc} for details. The notation with $\eta =
M^2 V''/V$, used e.g. in \cite{Lyth:1999xn}, simply corresponds to the
substitution $\alpha \rightarrow\alpha + 1$.  The 3 classes of SR
models are {\it small fields} $(\alpha < -1)$, {\it large fields} $(-1
< \alpha < 1)$, and {\it hybrid models} $(1 < \alpha)$.  
One finds \beq r = 2 \kappa \epsilon~, \, \, \, \, \, \dlnk = -2 \xi^2 
+ 8\epsilon^2 (2 \alpha -1) \, \, \, \, \, \mbox{and} \, \, \, \, \, n-1
= 2 \epsilon (\alpha-2)~.
\label{appendixEQ1}
\eeq


\end{document}